\documentstyle[preprint,eqsecnum,tighten,floats,epsfig,12pt,prl,aps]{revtex}

\begin{document}
\title{A New Upper Limit for the\\
Tau-Neutrino Magnetic Moment}
\author{
DONUT Collaboration\\
R. Schwienhorst$^1$, R. Rusack$^1$, D. Ciampa$^1$, C. Erickson$^1$, \\
M. Graham$^1$, K. Heller$^1$, J. Sielaff$^1$, J. Trammell$^1$, \\
J. Wilcox$^1$ K. Kodama$^2$, N. Ushida$^2$, C. Andreopoulos$^3$, \\
N. Saoulidou$^3$, G. Tzanakos$^3$, P. Yager$^4$, B. Baller$^5$, \\
D. Boehnlein$^5$,W. Freeman$^5$, B. Lundberg$^5$, J. Morfin$^5$, \\
R. Rameika$^5$, J.C. Yun$^5$, J.S. Song$^6$, C.S. Yoon$^6$, \\
S.H.Chung$^6$, P. Berghaus$^7$, M. Kubantsev$^7$, N.W. Reay$^7$, \\
R. Sidwell$^7$, N. Stanton$^7$, S. Yoshida$^7$, S. Aoki$^8$, T. Hara$^8$, \\
J.T. Rhee$^9$, K. Hoshino$^{10}$, H. Jiko$^{10}$, M. Miyanishi$^{10}$, \\
M. Komatsu$^{10}$, M. Nakamura$^{10}$, T. Nakano$^{10}$, K. Niwa$^{10}$, \\
N. Nonaka$^{10}$, K. Okada$^{10}$, O. Sato$^{10}$, T. Akdogan$^{11}$, \\
V. Paolone$^{11}$, C. Rosenfeld$^{12}$, A. Kulik$^{11,12}$, \\
T. Kafka$^{13}$, W. Oliver$^{13}$, T. Patzak$^{13}$, J. Schneps$^{13}$\\
\footnotesize $^1$ {\em University of Minnesota, Minnesota }\\
\footnotesize $^2$ {\em Aichi University of Education, Kariya, Japan}\\ 
\footnotesize $^3$ {\em University of Athens, Athens 15771, Greece}\\
\footnotesize $^4$ {\em University of California/Davis, Davis, California}\\
\footnotesize $^5$ {\em Fermilab, Batavia, Illinois 60510}\\
\footnotesize $^6$ {\em Gyeongsang National University, Jinju 660-701, Korea}\\
\footnotesize $^7$ {\em Kansas State University, Manhattan, Kansas}\\
\footnotesize $^8$ {\em Kobe University, Kobe, Japan}\\
\footnotesize $^9$ {\em Kon-kuk University, Korea}\\
\footnotesize $^{10}$ {\em Nagoya University, Nagoya 464-8602, Japan}\\
\footnotesize $^{11}$ {\em University of Pittsburgh, Pittsburgh, Pennsylvania 15260}\\
\footnotesize $^{12}$ {\em University of South Carolina, Columbia, South Carolina}\\
\footnotesize $^{13}$ {\em Tufts University, Medford, Massachusetts 02155}
}
\date{\today}
\maketitle
\begin{abstract}
Using a prompt neutrino beam in which a $\nu_\tau$ component was
identified for the first time, the $\nu_\tau$ magnetic moment was
measured based on a search for an anomalous increase in the number of 
neutrino-electron interactions. One such event was observed when 2.3
were expected from background processes, giving an upper 90\%
confidence limit on $\mu_{\nu_\tau}$ of $3.9\times 10^{-7} \mu_B$.
\begin{tabbing}
PACS numbers 14.60.Lm, 14.60.St, 13.40.Em
\end{tabbing}
\end{abstract}

\narrowtext

\section{Introduction}
\label{sec:level1}
Magnetic moment measurements are an important tool for probing the 
fundamental structure of matter. Currently, precision measurements of 
the electron and muon magnetic moments are being used to probe
the structure of the vacuum to the highest precision.  
A non-zero magnetic moment for any neutrino would be a clear and unambiguous 
signal for physics beyond the standard model. 
We have used the data collected for the Fermilab experiment E872 (DONUT)
to perform a search for anomalous
electromagnetic interactions of the tau-neutrino that would be a 
signature for a magnetic moment.
This paper presents the first measurement of the $\nu_\tau$ 
magnetic moment using a neutrino beam in which the $\nu_\tau$'s have been 
positively identified.

The direct limits on $\nu_e$ and $\nu_\mu$ magnetic moments are 
$\mu_{\nu_e} < 1.8 \times 10^{-10}\mu_B$ \cite{derbin94}
and $\mu_{\nu_\mu} < 7.4 \times 10^{-10}\mu_B$ \cite{krakauer90} and 
efforts are currently being made to extend
these limits by another order of magnitude \cite{trofimov,beda98}.
Because no direct measurements of the $\nu_\tau$ component
of a neutrino beam has been made until now, there has been no 
comparable measurement of its magnetic moment. 
In the CERN experiment WA66\cite{wa66}, a
limit of $\mu_{\nu_\tau} < 5.4\times10^{-7}\mu_B$  was based on an
assumed flux of $\nu_\tau$'s in the neutrino beam \cite{bebc87}.  
Elsewhere, indirect limits were 
derived from the duration of the supernova explosion SN1987A,
giving a limit of $10^{-11}\mu_B$ for all neutrino flavors
under the assumption that all neutrino flavors are equally produced 
in the supernova explosion. 
 
More recently it has been argued \cite{gninenko99} 
that the oscillation of atmospheric $\nu_\mu$
to $\nu_\tau$ would yield a limit of $\mu_{\nu_\tau} < 1.3\times10^{-7}\mu_B$.
However, it has been pointed out that this
value should be interpreted as a limit on the initial state
$\nu_\mu$ instead \cite{beacom}.

It used to be conventional to associate the properties of the neutrinos with
their weak eigenstates $\nu_e$, $\nu_\mu$, and $\nu_\tau$. 
The discovery of neutrino oscillations \cite{superk,soudan2} implies
that the electromagnetic properties of neutrinos should be 
associated with their mass eigenstates as opposed to their weak 
eigenstates \cite{beacom}. Since the parameters of the
mixing matrix are as yet undetermined, and there remains
the possibility that neutrino mixing is further complicated 
by an additional sterile neutrino \cite{bilenky98}, the composition 
of any neutrino beam in terms of the mass eigenstates can at 
this stage not be determined.
Until the oscillation scenario is fully understood, the
electromagnetic properties should therefore be characterized by the initial
neutrino flavor. This description is independent of any oscillation
assumptions and allows for an extraction of the mass eigenstate
magnetic moments when all of the mixing parameters are known \cite{beacom}.

In our measurement of the tau-neutrino magnetic moment we searched for an
anomalous increase in the elastic neutrino-electron cross-section
above the value predicted by the Standard Model. 
In the Standard Model neutrinos interact with electrons through $\text{Z}^0$
exchange, and a magnetic moment $\mu_\nu$ adds an extra
component due to photon exchange. 
The cross-section for a neutrino interacting via its magnetic moment with
an electron is given in the high-energy limit by
\begin{equation}
\label{eq1}
\frac{d\sigma_\mu}{dT_e} = \frac{\mu_\nu^2}{\mu_B^2}
 \frac{\pi \alpha^2}{m_e^2}\left( \frac{1}{T_e} - \frac{1}{E_\nu} \right),
\end{equation}
where $T_e$ is the energy of the scattered electron in the laboratory
frame \cite{magmomsigma}. The total magnetic moment scattering cross
section is obtained by integrating over $T_e$. The lower integration
limit is given by the experimental threshold for low-energy electrons 
(0.1~GeV in this analysis).
Since the neutrino undergoes a spin-flip
when a photon is exchanged, there is no interference with the Standard
Model process, and the total neutrino-electron scattering cross-section 
is just given by the sum of the two contributions.

Kinematic constraints \cite{radel93} limit the angle between the
incoming neutrino and the scattered electron in the laboratory frame to be 
\begin{equation}
\label{eq3}
\theta_{\nu-e}^2 < 2m_e / E_e 
\end{equation}
and for electron energies in excess of about 1~GeV, $\theta_{\nu-e}$
is less than 30~mrad. This angular constraint can be used as a clear signal 
to select neutrino-electron scattering events from the background of
$\nu_e$-nucleon charged-current events in which the electron is 
produced at a much larger angle.

\section{The apparatus}

The Fermilab experiment E872 (DONUT) took data in 1997 and
$\nu_\tau \text{-N}$ charged current interactions were observed in
nuclear emulsion \cite{donutdiscovery}. 
Having established the existence of $\nu_\tau$ in the $\nu$ beam and 
measuring the ratio of $\nu_e$, $\nu_\mu$ and $\nu_\tau$ using 
charged current interactions, we searched for single electrons 
arising from $\nu$ interactions, the signature of a $\nu$ magnetic moment.
The apparatus is described in detail elsewhere \cite{donutnim}; 
only the components central to this analysis will be discussed here.

The experiment consisted of three essential parts: 
a $\nu_\tau$ enriched neutrino beam produced in a beam-dump target, 
a neutrino target, and a spectrometer with electron and muon identification. 
A ``prompt'' neutrino beam
was produced by a beam of 800~GeV protons incident on a tungsten target.
The target length and material were chosen so that most of the 
long-lived secondary particles would interact in the target 
before decaying, 
while the short-lived particles would decay before interacting.
Hence the main contribution to the
high-energy neutrino flux came from the decay of the short-lived 
D-mesons, with $\nu_e$'s and $\nu_\mu$'s being produced in their 
semi-leptonic decays, and the $\nu_\tau$'s primarily from the decay channel
$D_s\rightarrow \tau \overline{\nu}_\tau$, $\tau\rightarrow \nu_\tau + X$. 
Downstream of the tungsten target there was 30~m of absorber material
and two sweeping magnets to remove all but the neutrinos from the beam.
>From our flux calculations using the LEPTO \cite{lepto} particle
generator and a GEANT \cite{geant} simulation, we estimate that in the sample
of located events,the neutrino interactions would be 47\% $\nu_\mu$ 
charged current events, 27\% $\nu_e$ charged current events, 
5\% $\nu_\tau$ charged current
events, and 21\% neutral current events. This composition was confirmed by 
our measurements of charged current interactions \cite{donutdiscovery}.

\begin{figure}
\epsfig{file=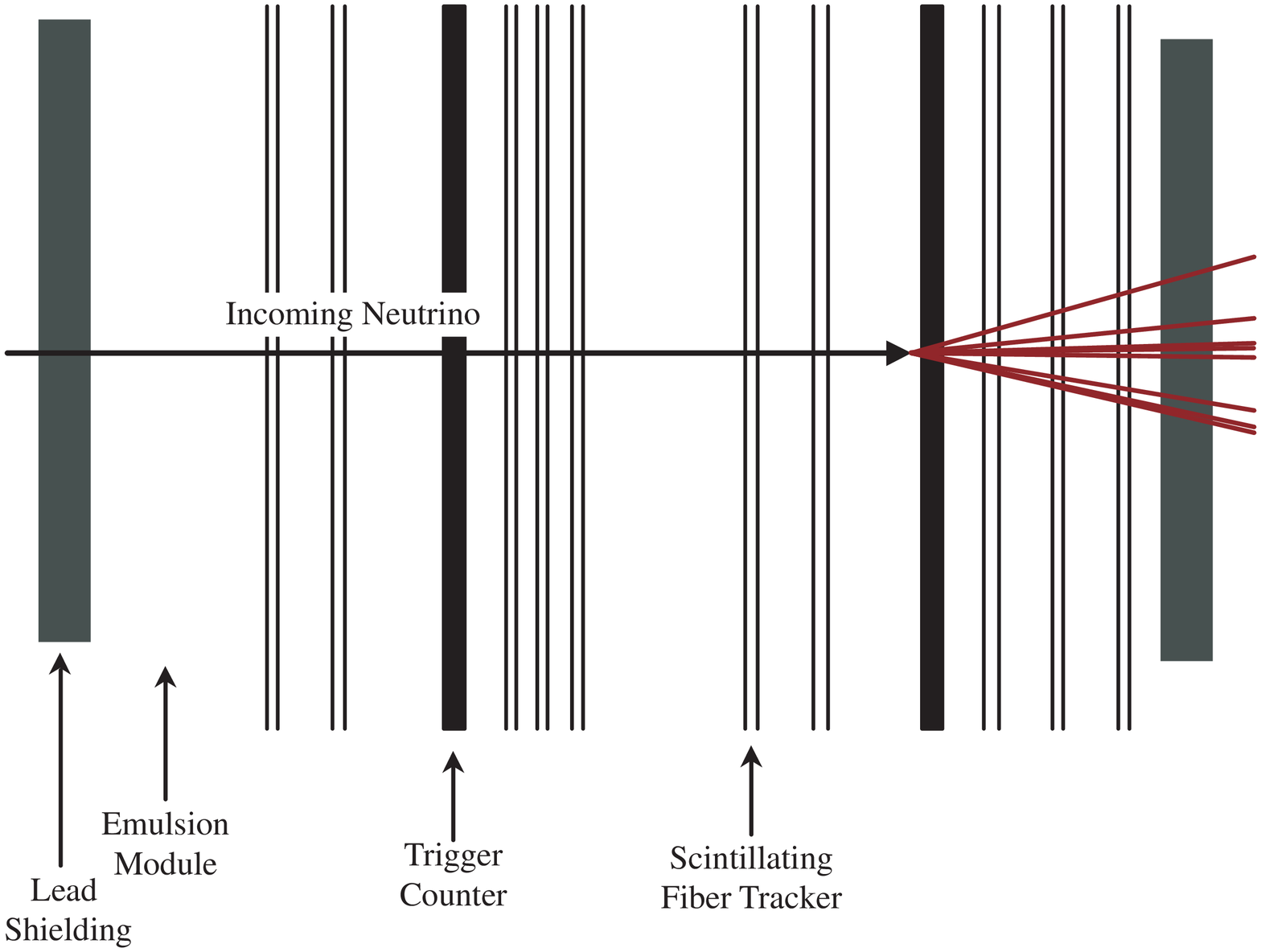,height=9cm}
\caption{Target region view of the selected magnetic moment candidate
event. The neutrino is incident from the left. }
\label{fig1}
\end{figure}

Behind the absorber was the neutrino target region, shown in figure
\ref{fig1}, consisting of four steel and emulsion modules that were interleaved
with planes of 0.5~mm diameter scintillating fibers. 
The entire region was surrounded 
by lead shielding, 20~mm thick upstream and 6~mm thick downstream.  
The total target mass, including the upstream lead, was 554~kg. 
In front of the lead shielding was a wall of scintillator counters used 
in the trigger to veto interactions upstream of
the target region. 

The spectrometer consisted of the scintillating fiber tracker 
in the target region, drift chambers upstream and downstream of a 
wide-aperture analysis magnet, and a lead glass array
with a muon identification system behind it. The data acquisition 
system was triggered by a coincidence between a plane of 
scintillator counters just downstream of the target region 
and at least one of two planes of scintillators
inside the region, together with no signal in the upstream veto wall.

In the search for anomalous $\nu \text{-e}$ interactions, the 
steel and emulsion target modules interleaved with
the fiber tracker served as a 
high-granularity sampling calorimeter that allowed the identification 
and location of electromagnetic showers.

\section{Event Selection}

In the six-month run of the experiment $4.0 \times 10^6$ events were
recorded for analysis and $2.0 \times 10^5$ of these 
had two or more tracks in the target region. 
Candidate neutrino-electron scattering events were selected from these
events if the following requirements were met:
there were no identified muons associated with the event;
there was a signal in the trigger hodoscope directly downstream of 
the reconstructed vertex;
the reconstructed tracks had an angle of less than
100~mrad with respect to the neutrino direction, 
and the reconstructed vertex was in the fiducial volume (in a target
module and no more than 0.24~m in the horizontal direction from the
center of a module).

These cuts removed events triggered by high-momentum muons that 
interacted in the material surrounding the target and its 
support system sending
secondary particles into the target area.

After these cuts were applied, events with a reconstructed vertex that
was consistent with a neutrino interaction were selected in a visual
scan, leaving 285 events in the data compared to 186 expected from a 
Monte Carlo simulation. The excess in the data was caused
by the interaction of low-energy neutrons or photons in the most downstream
target module. These events were removed by requiring a signal of at 
least 2~GeV in the lead glass calorimeter if the interaction vertex
was in the most downstream target module. This cut reduced the number
of events to 100 in the data and 95 in the Monte Carlo.

The remaining $\nu\text{-N}$ scattering background events were removed 
if there were any tracks in the event not associated with an electron
or if secondary particles from the nuclear breakup were identified 
upstream of the interaction vertex.
Events were also rejected if any reconstructed track was identified 
in the muon ID system, if the ratio of the measured momentum of a track to the 
signal in the electromagnetic calorimeter was less than 0.5, if a
track crossed more than two radiation lengths of target material
without initiating a shower, or if the track left a 
sequence of large energy deposits in neighboring layers of the 
scintillating fiber tracker.
Once these tests were applied 13 events remained. 

As a final test, events were selected if  
there was an identified electron at an angle of less than 30~mrad 
with respect to the neutrino direction and if there were no
reconstructed large-angle
tracks with angles in excess of 500~mrad. 
This left one event, shown in figure \ref{fig1}, 
an interaction in the most downstream target module.

\section{Data Analysis}

The expected number of events from Standard Model processes was
determined by a GEANT Monte Carlo simulation of the apparatus. The
same program was
used to find the selection efficiency for magnetic moment interactions. 
The accuracy of the simulation was tested by comparing data and 
simulation for two different sets of well-understood and easily 
identifiable events. The first set of control events consisted 200
$\nu_\mu$ charged-current interactions containing an identified muon. 

The second set consisted of 150 straight-through muons collected for
calibration and alignment that scattered off an electron in the target
region which in turn produced an electromagnetic shower.
The electron
energy spectrum of these ``knock-on'' electrons drops as $1/T_e^2$, 
while for magnetic-moment interactions it depends on 
$1/T_e$ (see equation \ref{eq1}).
Though not identical, the similarity of the spectrum allows this
control set to be used to quantify the effect of each selection cut 
on electromagnetic showers.

As a test of our simulation each cut applied to the data 
was also applied to both sets of control events. No
significant difference between data and Monte Carlo was found with 
each cut removing the same fraction of events from both data and 
Monte Carlo for both control sets within statistical uncertainty.  

The Monte Carlo was also used to estimate the overall detection efficiency
for electron events. As the event trigger required
at least one hit in the most downstream of the trigger planes (T3), the
shower from electrons produced in the neutrino interaction had to pass
through two radiation lengths for each emulsion module.
The acceptance was further limited by a 6 mm thick lead shield that was placed 
between the target region and T3. 
The overall trigger efficiency for events where neutrinos interact 
electromagnetically with electrons was $(26.8\pm1.4)$\%,
ranging from 10\% for interactions in the upstream lead wall to
50\% for interactions in the most downstream module and,
after the neutrino interaction selection cuts $(15.2\pm0.8)$\% 
of this type of events remained. By further requiring an  
identification of an electromagnetic shower in the event, the 
overall selection efficiency was reduced to $(9.0\pm0.6)$\%.

\section{Flux calculation}
The fixed target D-meson production cross sections using a proton
beam are $\sigma_p(D^\pm)=(11.3 \pm 2.2)\mu b/\text{nucleon}$, 
$\sigma_p(D^0)=(28.0 \pm 2.5)\mu b/\text{nucleon}$, and 
$\sigma_p(D_s)=(5.2 \pm 0.8)\mu b/\text{nucleon}$ \cite{donutnubeam}. 
All cross sections were
found to increase linearly with the atomic mass. For comparison, the 
previous CERN experiment used a value of $\sigma_p(D_s)=2.6\mu
b/\text{nucleon}$ at a beam energy of 400~GeV\cite{wa66}, 
for which this calculation yields a value of $2.4\mu b/\text{nucleon}$.

The differential cross section for D production was modeled by
\begin{equation}
\label{sigmadf}
\frac{d^2\sigma}{dp_t^2 dx_F} \propto (1-|x_F|)^n \exp(-b p_t^2)
\end{equation}
with values of $n=(7.4 \pm 0.6)$ and $b=(0.94 \pm 0.06)\text{~GeV}^{-2}$ \cite{donutnubeam}.

The branching ratio for the decay $D_s\rightarrow \tau
\overline{\nu}_\tau$ is $(6.3 \pm 0.5)\%$, based on recent  
measurements of the decay\cite{cleods}. 
The resulting $\nu_\tau$ flux is $(2.1 \pm 0.3)
\times10^{-5}\nu_\tau m^{-2} \text{POT}^{-1}$.

\section{Results and Discussion}

The single event surviving all the cuts occurred in the most downstream module and
produced a narrow shower of particles with a reconstructed track at
its center that had an angle of $(10 \pm 5)$~mrad with respect to the neutrino
direction. The total recorded
signal in the calorimeter was 20.0~GeV, of which 16.8~GeV were associated with
the central part of the electromagnetic shower. For this
energy, the electron angle should be less than 7~mrad according to
Eq. (\ref{eq3}).
The selected event could be a quasi-elastic $\nu_e\text{-N}$
interaction with a relatively low-energy neutrino. 

The expected background rate due to neutrino-nucleon scattering was 
2.3 events and other processes, such as neutral current neutrino-electron scattering, 
are expected to contribute less than 0.05 events. 
Since we observed one event when the expected background rate was 2.3
events, we conclude that we did not observe any signal events, and therefore
the measured magnetic moment is zero. A statistical analysis based on the
Feldman-Cousins method \cite{feldcous} yields a 90\% upper confidence 
limit of 2.3 events in the signal. 

To convert this limit in the number of signal events to a limit on 
the tau-neutrino magnetic moment, Eq. \ref{eq1} is integrated 
numerically, taking into account the calculated $\nu_\tau$ energy spectrum. 
This yields
\begin{equation}
\label{eq5}
\sigma_{tot}^\mu =  \frac{\mu_\nu^2}{\mu_B^2} \times 1.8\times 10^{-28} m^2.
\end{equation}

In the experiment a total of  $3.56 \times 10^{17}$ protons on target were recorded, and
the average target mass during the run was 554~kg, corresponding to
$1.7\times10^{29}$ target electrons, hence the number of expected events for
a given magnetic moment, given by the product of the flux, cross
section, and number of scattering centers, is:
\begin{equation}
\label{eq6}
n_{\text{events}} =  \frac{\mu_\nu^2}{\mu_B^2} \times 1.5\times 10^{13}.
\end{equation}

This gives an upper limit for the magnetic moment of 
$\mu_{\nu} < 3.9 \times 10^{-7} \mu_B$. 
This should be compared to the experimental sensitivity of 
$4.9 \times 10^{-7} \mu_B$, which is the limit that would be obtained had 
we observed as many events as predicted for the background. 

This analysis is flavor-blind and the limit applies in
principal to the sum over all neutrino flavors. However, more
stringent limits have been determined by other experiments for $\nu_e$ 
and $\nu_\mu$. Assuming a magnetic moment at the current limit, 
their contribution would be less than $10^{-4}$ events. The limit is
therefore interpreted as a new upper limit on $\nu_\tau$.

While systematic uncertainties were not included in this analysis
as they are not part of the Feldman-Cousins method,
they would only change the result slightly, since the uncertainty due 
Poisson statistics completely dominates the estimate of the limit. 
Contributions to
the systematic uncertainty come from the neutrino flux calculation (15\%),
the total proton flux (15\%), and the number of generated Monte Carlo
events (5\%). In addition to the statistical analysis based on the 
Feldman-Cousins method, 
a Bayesian analysis was performed including all these systematic
uncertainties \cite{dagostini} using a flat prior 
distribution in the magnetic moment. This yields a 90\% confidence limit of 
$\mu_{\nu_\tau} < 3.5 \times 10^{-7} \mu_B$. We include this result for 
comparison. However, only the upper limit derived with the Feldman-Cousins method
should be quoted. 

\section{Conclusions}
The new upper limit for the tau-neutrino magnetic moment of $3.9
\times 10^{-7} \mu_B$ is an improvement over the previous limit \cite{wa66}. 
Moreover, this is the first experiment to directly observe 
$\nu_\tau$ charged-current interactions and therefore be certain to 
have a $\nu_\tau$ component in the neutrino beam.

The new limit is still three orders of magnitude above the limits for
$\nu_e$ and $\nu_\mu$, and it dominates when extracting limits for the
mass eigenstates from the current limits for the flavor eigenstates.
Improving the limit on $\mu_{\nu_\tau}$ would require a $\nu_\tau$ beam
that is comparable in total flux to previous $\nu_e$ and $\nu_\mu$ beams.

\acknowledgments
We gratefully acknowledge the ingenuity and support given to us by the staffs at
Fermilab staff and at the collaboration universities. 
This work is supported by the 
the General Secretariat of Research and Technology of Greece, 
the Japan Society for the Promotion of Science,
the Japan-US Cooperative Research Program for High Energy Physics,
the Ministry of Education, Science and Culture of Japan,
the Korea Research Foundation Grant,
and United States Department of Energy.

\end{document}